\documentclass{article}
\usepackage{spconf,amsmath,graphicx}
\usepackage{booktabs, multirow} 
\usepackage{soul}
\usepackage[table]{xcolor} 
\usepackage{changepage,threeparttable} 
\usepackage[ruled,vlined,linesnumbered]{algorithm2e}

\title{Maximum-Entropy Adversarial Audio Augmentation for Keyword Spotting}
\name{Zuzhao Ye$^{1,2}$, Gregory Ciccarelli$^{1}$, Brian Kulis$^{1,3}$}
\address{
  $^1$Amazon Inc.\\
  $^2$University of California, Riverside\\
  $^3$Boston University}

\begin{document}

\onecolumn
\vspace*{\fill}
\noindent © 2024 IEEE. Personal use of this material is permitted. Permission from IEEE must be obtained for all other uses, in any current or future media, including reprinting/republishing this material for advertising or promotional purposes, creating new collective works, for resale or redistribution to servers or lists, or reuse of any copyrighted component of this work in other works.
\vspace*{\fill}
\twocolumn
\newpage

\maketitle

\begin{abstract}
Data augmentation is a key tool for improving the performance of deep networks, particularly when there is limited labeled data.  In some fields, such as computer vision, augmentation methods have been extensively studied; however, for speech and audio data, there are relatively fewer methods developed.  Using adversarial learning as a starting point, we develop a simple and effective augmentation strategy based on taking the gradient of the entropy of the outputs with respect to the inputs and then creating new data points by moving in the direction of the gradient to maximize the entropy.  We validate its efficacy on several keyword spotting tasks as well as standard audio benchmarks. Our method is straightforward to implement, offering greater computational efficiency than more complex adversarial schemes like GANs. Despite its simplicity, it proves robust and effective, especially when combined with the established SpecAugment technique, leading to enhanced performance.
\end{abstract}

\begin{figure*}[!t]
\centering
\includegraphics[width=0.90\textwidth]{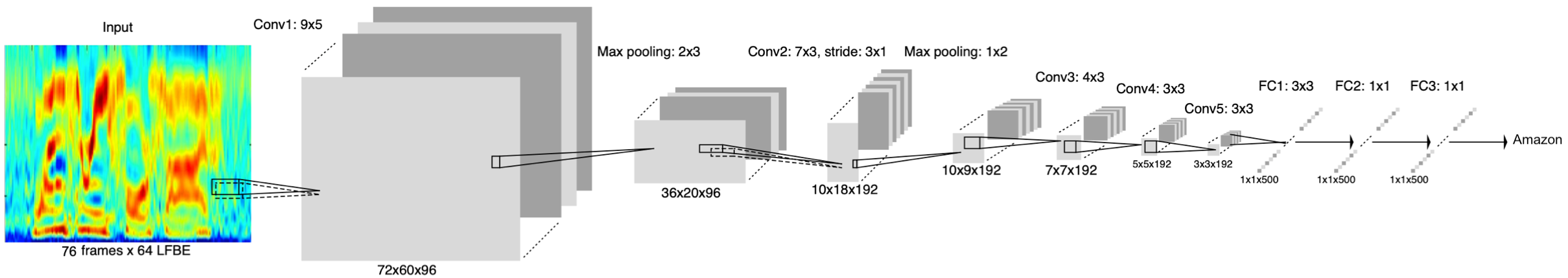}
\caption{CNN architecture used in the experiments~\cite{jose}.}
\label{fig:cnn_diagram.png}
\vspace{-3mm}
\end{figure*}

\begin{figure}[!t]
\includegraphics[width=.45\textwidth]{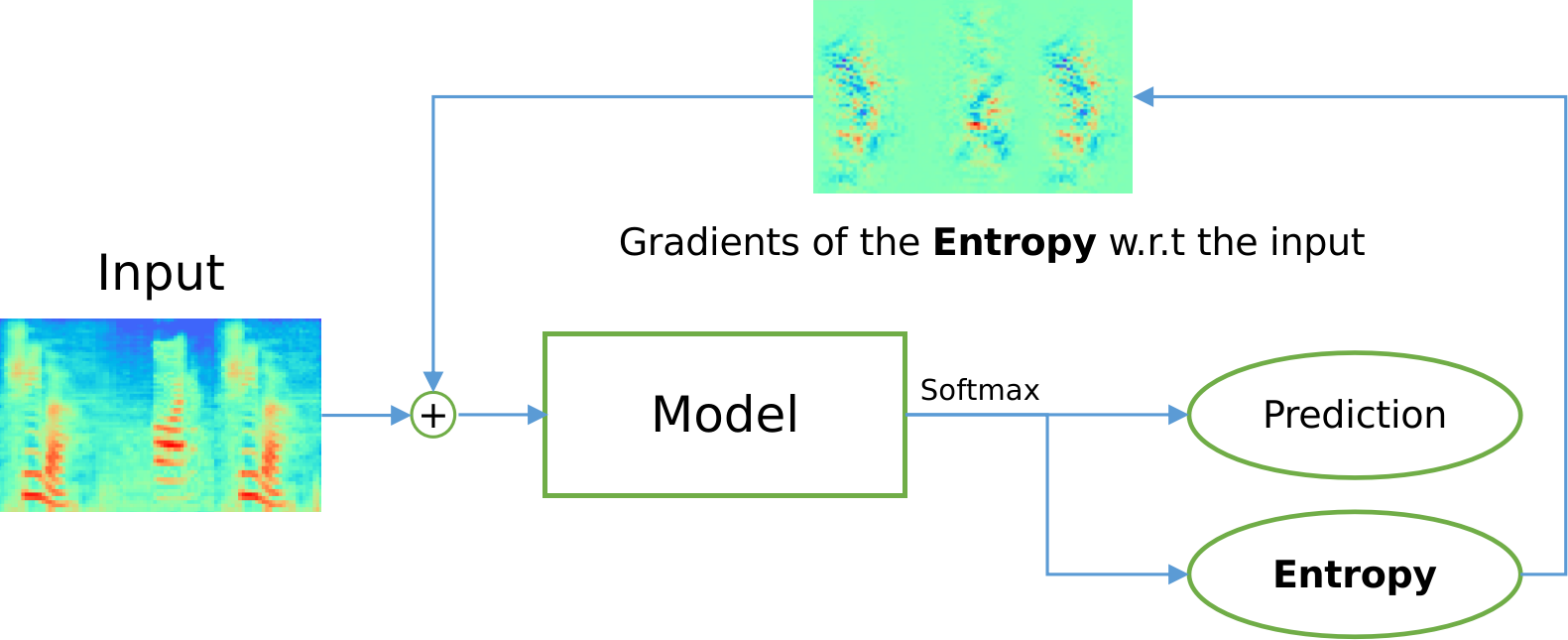}
\caption{Overview of adversarial training with entropy. An input data point is fed through the architecture, and then the entropy is calculated for the softmax output. The gradient of the entropy with respect to the input is used to perform a gradient ascent step in order to produce an augmentation.}
\label{fig:approach.png}
\vspace{-3mm}
\end{figure}

\vspace{-.3cm}
\section{Introduction}
\vspace{-.3cm}
Modern deep learning architectures have been remarkably powerful at solving difficult learning problems, but typically require huge amounts of labeled data in order to be effective. Recent research efforts have extensively explored solutions to mitigate this issue, including few-shot learning, domain adaptation, transfer learning, and data augmentation~\cite{zhang2021dive}.
Data augmentation in particular has been an extensively-studied problem in the computer vision domain~\cite{augmentation_survey}.  In that setting, strategies such as rotations, mirror imaging, and cropping have been developed to augment visual data. It is standard practice to utilize multiple augmentation schemes for many learning tasks in vision~\cite{fixmatch}.

Unlike in computer vision, in the audio domain, relatively less attention has been paid to augmentation.  The current gold standard for augmentation in the spectrogram domain is SpecAugment~\cite{specaugment}, which applies various transformations to the spectral representation of the audio (e.g., time and frequency masking) to create new data samples for training. While SpecAugment has been shown to be effective in many settings, a drawback is the need for careful tuning of hyperparameters, such as masking ratios, to avoid potential harm to the model. 


Our motivation in this paper is to develop a simple, robust, and efficient augmentation strategy for audio data.  To be computationally simple, we  eschew methods that involve complex optimization; for instance, methods such as generative adversarial networks can be used to augment the data, but require expensive training prior to use.  Our approach is inspired by the method of adversarial learning~\cite{goodfellow_adv}, which was proposed for the task of building classifiers that are robust to adversarial attacks.  Our aims are different from theirs; we are interested in augmentation for training with limited data, but we borrow their idea of constructing examples by computing gradients with respect to the input data.  We also note that there are some existing adversarial augmentation schemes~\cite{zhao2020maximum,hsu_aaai} but these approaches involve solving expensive min-max optimizations, counter to the goal of our approach.

Our method works by taking the softmax output of the classifier, treating it as a probability distribution, computing its entropy, and then backpropagating to compute the gradient of the entropy with respect to the input data.  Then we apply gradient ascent to construct new data points that increase the entropy of the classifier.  The idea of maximizing entropy for limited supervision is well-established in various fields~\cite{saito,dubey}. Computationally, this requires a second back-propagation pass, thus roughly doubling the time to do back-propagation: one pass to perform the augmentation, and another pass to   use the augmented data to update the model.

We focus on the problem of keyword spotting. In this setting, the goal is to identify whether a piece of audio contains a particular keyword or not. Data augmentation is particularly useful given the number of challenging keyword spotting (a.k.a. wakeword detection) tasks that arise. For Amazon in particular, different models are trained in varied scenarios, ranging from different languages and locales to different device types to different keywords. As a result, training separate models for each scenario presents engineers with potential data scarcity for supervised training methods.

We show on a series of Amazon keyword spotting tasks that: a) performance increases with our proposed augmentation scheme, yielding comparable results to SpecAugment, and b) when combined with SpecAugment, we produce better performance than with any single augmentation strategy. We also experiment with our technique on a series of standard public audio benchmark data sets, demonstrating the robustness, efficacy, and computational efficiency of our method. Thus, we conclude that our approach may provide a benefit for use as an augmentation strategy, and warrants further exploration for other speech and audio tasks.

\vspace{-.3cm}
\section{Background and Related Work}
\vspace{-.3cm}

\textbf{Data Augmentation}: Data augmentation is an important problem in various tasks in audio.  Examples of successful applications include singing voice detection~\cite{voice_detection}, recognition of accented speech~\cite{accented}, animal audio classification~\cite{animal_classification}, and musical instrument recognition~\cite{instrument_recognition}.  In general, data augmentation for audio has been studied in the raw audio domain as well as the spectrogram domain.  In the raw audio domain, common methods include time shifting, time stretching, pitch scaling, noise addition, impulse response addition, filters, polarity inversion, and random gain.  In the time-frequency domain with spectrograms, some similar ideas have been pursued (e.g., pitch shifting, time shifting, or time stretching), but the most widely adopted augmentation method is SpecAugment~\cite{specaugment}.  This method works by applying time masking and/or frequency masking to the spectrogram, and has been particularly effective for ASR tasks.

Besides SpecAugment, some other recent methods for augmentation have been proposed.  In \cite{emotion_recognition}, they  train a generative adversarial network in order to generate artificial data.  Another recent idea is~\cite{animal_classification}, in which combinations of techniques in the raw and spectrogram domains are considered.  In \cite{specmix}, the authors propose a method called SpecMix which is based on SpecAugment, but with mixed labels.

In contrast to existing work, we focus mainly on a technique that can be applied to both the raw and spectrogram domains, although we focus experimentally on the spectrogram domain.  Furthermore, unlike methods such as GANs, our technique is extremely simple. The study \cite{zhao2020maximum} resonates with our idea, but it applies the maximum entropy adjustment merely in the few initial training steps, producing a training set many times larger than the original and significantly compromising training efficiency.  It also involves a more complex min-max optimization procedure.  The method of~\cite{hsu_aaai} is also superficially related to ours but is designed for the unsupervised setting, making a direct comparison impossible.

\textbf{Adversarial Training:} Adversarial training attempts to train models to be robust to adversarial attacks.  A common way to train methods to be robust is to construct adversarial examples during training that ensure that the resulting model can handle such adversarial examples.  While this is a different motivation from data augmentation, we use this approach to inspire our method. In the Fast Gradient Sign Method (FGSM) method~\cite{goodfellow_adv}, a simple adversarial scheme is suggested.  Here, an adversarial example is constructed via:
\begin{displaymath}
x_{adv} = x + \epsilon \cdot \mbox{sign}(\nabla_x \ell(x, y, \theta)),
\end{displaymath}
where $\ell$ is the loss function of the model, $x$ is the training data, $y$ is the supervision, and $\theta$ are the parameters of the underlying model.  Follow-up work has considered variants of this approach as well as faster training strategies to compute and use the adversarial examples with only a single backpropagation pass~\cite{adversarial_free}.

We note that our approach differs in that our goal is data augmentation, not adversarial training.  Indeed, adversarial training with the loss function is known to improve robustness to adversarial attacks but often yields worse accuracy. Therefore, it is worth emphasizing that we perform the adversarial on entropy rather than the loss function.

\vspace{-.3cm}
\section{Approach}
\vspace{-.3cm}
Our goal is to develop a novel technique for audio augmentation that maintains simplicity, is computationally inexpensive, and performs well. To that end, we develop a simple adversarial-style training approach but, unlike previous adversarial training methods, our approach is focused on augmentation rather than preventing adversarial attacks.

Suppose we have a network architecture that is designed for certain audio tasks (e.g. keyword spotting, classification, generation). We are agnostic to the task but assume a classification setting in which the output at the last layer of the network is a softmax over class labels. This output can be interpreted as a probability distribution, allowing us to calculate its entropy. During training, we can generate augmentations as follows: We pass an input $x$ through the network and compute the entropy right after the softmax output. We take the gradient of the entropy with respect to $x$, and then move $x$ in the direction of the gradient (gradient ascent) to maximize the entropy. This yields a new data point $x_{aug}$, which is used for the subsequent training.

Figure~\ref{fig:approach.png} illustrates our augmentation approach, which we call Adversarial Training with Entropy (ATE).  In particular, we will utilize the following equation for constructing an augmented data point. Given a binary softmax (sigmoid) output $p$, we compute the entropy as $E = -[p \log p + (1-p) \log (1-p)]$ (Note this is different from cross entropy).  Then we construct the augmentation via
\begin{displaymath}
x_{aug} = x + \mbox{clip}(\nabla_x E(x, y, \theta), -\epsilon, \epsilon),
\end{displaymath}
where $x$ is the input data, $y$, is the supervised class label, $\theta$ are the model parameters, and the clip operation clips the gradient values between $-\epsilon$ and $\epsilon$. For multiclass classification problems, the entropy is simply $E=-\sum_{i=1}^{N}p_i \log p_i$, where $p_i$ is the softmax output for class $i$ and $N$ is the number of classes. Note that this procedure requires two forward and backward passes: one to construct the augmentations, and another to update the weights based on the augmented data. The overall training procedure is summarized in Algorithm \ref{algo:ATE}.
\vspace{-0.2cm}

\begin{algorithm}[h!]
\SetAlgoLined

\textbf{Input:} Dataset $\mathcal{D}$ and initial network with weight $\theta^0$\;
\textbf{Output:} Trained weight $\theta$\;

\For{step = 1:n}{
Sample a batch of data $(x,y)$ from $\mathcal{D}$\;
\eIf{random(0,1) $< P^{aug}$}{
    $x_{aug} \leftarrow x + \mbox{clip}(\nabla_x E(x, y, \theta), -\epsilon, \epsilon)$\;
    $\theta \leftarrow \theta - \alpha\nabla_{\theta} \ell(x_{aug}, y, \theta)$
    }{
    $\theta \leftarrow \theta - \alpha\nabla_{\theta} \ell(x, y, \theta)$
    }
}
\caption{Adversarial Training with Entropy}
\label{algo:ATE}
\end{algorithm}
\vspace{-0.3cm}

Our use of entropy followed by gradient ascent is chosen based on a plethora of work demonstrating the effectiveness of maximum entropy in a variety of unsupervised, semi-supervised, and supervised settings~\cite{saito,dubey}.  The principle of maximum entropy suggests choosing a distribution with the highest entropy to represent the current state of knowledge.  Ref~\cite{dubey} proposes a maximum entropy classifier in the context of fine-grained classification, where maximum entropy is applied to the prediction vectors.  We apply a similar idea here; by creating an augmented data point that increases the entropy of the prediction vector, we hope to provide a useful augmentation that follows these guiding principles.

\vspace{-.3cm}
\section{Experiments}
\vspace{-.3cm}
\subsection{Data}
\vspace{-.2cm}

We conduct experiments on Amazon proprietary data sets and public data sets; both are annotated. For the former, we select keywords (i.e. wakewords) ``Computer,'' ``Amazon,'' and ``Echo'' in the India locale, where there is less annotated data. The customer data is de-identified. For each keyword, we consider a small data set with 10,000 data points, and a larger one with 100,000 data points. For the public data sets, we select ESC-50 ~\cite{ESC50}, UrbanSound8K (US8K)~\cite{UrbanSound8K}, and Speech Commands version 2 (SCV2)~\cite{SpeechCommands}. These are well-known datasets and each has 2,000, 8,000, and 105,000 data points, respectively. We first convert all audio into the spectrogram domain via 64-dimensional log Mel-filterbank energy (LFBE) features, computed over 25 ms frames with a 10 ms shift.

\vspace{-.2cm}
\subsection{Model and Training Details}
\vspace{-.2cm}
Figure~\ref{fig:cnn_diagram.png} shows the model following previous work~\cite{jose,sainath}. This model contains 5 convolutional layers and 3 fully connected layers, resulting in a model with approximately 2 million parameters. This model directly fits the Amazon data sets. For the public data sets, the kernel sizes are slightly adjusted to maintain integer outputs of each convolutional layer given different input dimensions in the time domain.

We train the model with Adam~\cite{Kingma2014AdamAM}. For Amazon data sets, we train for 80 epochs and use a batch size of 128.  The learning rate is set to 0.001, which we multiply by 0.3 every 8 epochs if the validation loss does not decrease. For public data sets, we use batch sizes of 45, 64, and 128 for ESC-50, US8K, and SCV2, respectively. These selections are based on a preliminary parametric study that yields the best results. Each public data set is trained for 300 epochs and the model checkpoint with the best validation accuracy is saved. The clipping threshold of ATE ($\epsilon$) is set at one standard deviation of the training data and the probability for performing augmentation ($P^{aug}$) is set at 0.5 for all experiments. 

\vspace{-.3cm}
\begin{table}[h]\centering
\caption{Comparison of augmentation methods on selected Amazon data sets (FAR at fixed FRR).}\label{tab:results}
\scriptsize
\begin{tabular}{lcrrrrrr}\toprule
Keyword &Size &\textbf{NoAug} &\textbf{SpecAug} &\textbf{ATE} &\textbf{S+A} &\textbf{A+S} \\\midrule
\multirow{2}{*}{\textbf{AMAZON}}   & 10k &0.059 &0.045 &0.052 &0.045 &\textbf{0.041} \\
                                   & 100k &0.037 &0.034 &0.034 &0.034 &\textbf{0.033} \\
\multirow{2}{*}{\textbf{ECHO}}     & 10k &0.110 &0.093 &0.099 &0.09 &\textbf{0.085} \\
                                   & 100k &0.055 &0.057 &\textbf{0.054} &0.056 &\textbf{0.054} \\
\multirow{2}{*}{\textbf{COMPUTER}} & 10k &0.036 &0.022 &0.024 &0.021 &\textbf{0.019} \\
                                   & 100k &0.018 &0.015 &0.013 &0.014 &\textbf{0.011} \\
\bottomrule
\end{tabular}
\end{table}

\begin{table*}[!htp]\centering
\caption{Comparison of augmentation methods on selected public data sets (Accuracy and training time).}\label{tab:results-public}
\scriptsize
\begin{tabular}{lrrrrrrrrr}\toprule
Dataset &\textbf{NoAug} &\textbf{SpecAug (S)} &\textbf{SpecMix (SM)} &\textbf{ME-ADA} &\textbf{ATE (A)} &\textbf{S+A} &\textbf{A+S} &\textbf{SM+A} &\textbf{A+SM} \\\midrule

\textbf{ESC-50} &0.566$\pm$0.017 &0.533$\pm$0.020 &0.541$\pm$0.028 & 0.558$\pm$0.021 &\textbf{0.581}$\pm$0.016 &0.535$\pm$0.017 &0.537$\pm$0.017 &0.540$\pm$0.044 &0.551$\pm$0.027 \\

\textbf{US8K} &0.742$\pm$0.042 &0.763$\pm$0.048 &0.760$\pm$0.046 & 0.760$\pm$0.031 &0.744$\pm$0.041 &0.766$\pm$0.049 &\textbf{0.767}$\pm$0.050 &0.757$\pm$0.046 &0.759$\pm$0.052 \\

\textbf{SCV2} &0.905$\pm$0.001 &0.888$\pm$0.019 &0.898$\pm$0.010 &\textbf{0.912}$\pm$0.003 &0.909$\pm$0.006 &0.887$\pm$0.011 &\textbf{0.912}$\pm$0.004 &0.884$\pm$0.017 & 0.881$\pm$0.002 \\

\toprule
\multicolumn{2}{l}{Training Time per Epoch} & \\\midrule

\textbf{SCV2} &16.8s &22.8s &23.9s &48.4s &21.5s &27.5s &27.5s &28.6s &28.6s\\

\bottomrule
\end{tabular}
\end{table*}

For evaluation, false accept rate (FAR) and false reject rate (FRR) are used following the practices for Amazon data sets, as they are binary classification tasks. We report the FAR at a fixed FRR (the precise FRR value is suppressed here given the proprietary nature of the data) to obtain easy-to-compare values. For public data sets, which are multiclass classification problems, we use accuracy as the measurement. For ESC-50 and US8K, we follow the standard 5-fold and 10-fold methods, respectively, to obtain the average accuracy. For SCV2, we follow the default train-validation-test split and also train three times for each augmentation method.

\vspace{-.3cm}
\subsection{Results}
\vspace{-.2cm}

Table~\ref{tab:results} presents results on the six Amazon data sets. In particular, we compare using no augmentation (NoAug) and SpecAugment \cite{specaugment}, but we also consider combining our approach with SpecAugment. This is done via two variants; one where SpecAugment is applied first (S+A), and the other where ATE is applied first (A+S). The approach with the best performance is the ATE followed by SpecAugment on all data sets. On the Echo-100k data set, the ATE method alone also matches this result. Our best results in terms of relative improvement occur on the smaller 10k data sets.  In particular, we obtain relative improvements over NoAug of 30.5, 22.7, and 47.2 percent on Amazon, Echo, and Computer, respectively, when using ATE followed by SpecAugment.

Table~\ref{tab:results-public} presents results on the three public data sets in terms of classification accuracy. We include additional baselines SpecMix~\cite{specmix} and ME-ADA~\cite{zhao2020maximum}, following the recommended settings from their original papers, for more comprehensive comparisons. Since SpecMix can be stacked with ATE just like SpecAugment, we also present the results of their combinations in different orders. We interpret the results in three different ways:
\vspace{-0.1cm}
\begin{itemize}
    \item \textbf{Robustness}. For the ESC-50, SCV2, and Amazon (ECHO at 100k) datasets, both SpecAugment and SpecMix tend to decrease accuracy. This raises concerns regarding the robustness of augmentation methods that predominantly utilize masking techniques. Contrastingly, ATE consistently demonstrates improvements across these datasets, underscoring its robustness. Additionally, ATE tends to exhibit less variance in accuracy as compared to SpecAugment and SpecMix, suggesting its stability.
    \vspace{-0.1cm}
    \item \textbf{Efficacy}. Although ATE as a standalone method is not consistently the top performer, its combination with SpecAugment (A+S) generally outperforms other baselines. Specifically, ATE achieves the best performance on the ESC-50 dataset and competitive performance on the SCV2 dataset. In the US8K dataset, while ATE does not outperform NoAug, the combination of ATE followed by SpecAugment does, indicating potential synergistic effects between the two methods. Interestingly, fusing ATE with SpecMix does not yield improvements. This observation prompts questions about the label-mixing procedure's potential incompatibility with ATE, which warrants further exploration.
    \vspace{-0.1cm}
    \item \textbf{Training Efficiency}. Analyzing the training times per epoch on the SCV2 data set, we find that both ATE and A+S strike a balance between performance and time cost, demonstrating only a moderate increase in the training duration. In contrast, ME-ADA, despite mirroring A+S's performance in this data set, demands considerably longer training time. This elongation stems from the inherent nature of ME-ADA's algorithm: every iteration of its minimax procedure doubles the dataset size. Consequently, with a default setting of two minimax operations, the training set quadruples, leading to a notably prolonged training phase.
\end{itemize}

Beyond the primary baselines, we tested other variants: random augmentation, adversarial augmentation using cross-entropy (typical of standard adversarial training), and adversarial augmentation targeting earlier network layers. These did not outperform SpecAugment, so we omitted their detailed reporting. We also explored text-to-speech augmentation, but like GAN-based augmentation, it demanded substantial additional machinery and did not show improvements.

\vspace{-.3cm}
\section{Conclusion}
\vspace{-.3cm}
We presented a novel data augmentation technique based on an adversarial-type training to maximize the classifier's output entropy. This approach yields a simple augmentation strategy that can be applied in a variety of settings and with a variety of input types. For our experiments, we focused on the keyword spotting problem with spectrogram inputs, and showed that our method yields robust improvements, particularly when paired with the state-of-the-art SpecAugment strategy. Additionally, through evaluations on widely-recognized public audio datasets, we validated the applicability and effectiveness of our approach in broader contexts.


\bibliographystyle{IEEEbib}
\bibliography{mybib}

\end{document}